\documentclass[twocolumn,prb,preprintnumbers,showpacs,amsmath,amssymb]{revtex4}
\usepackage{graphicx}
\usepackage{dcolumn}
\usepackage{txfonts}
\usepackage{xcolor}
\usepackage{amssymb}
\usepackage{amsmath}
\usepackage{latexsym}
\usepackage{epsfig}
\usepackage{amsbsy}
\usepackage{array}
\usepackage{tabularx}

\begin{document}

\title [] {The three-dimensional topological
Dirac semimetal Na$_3$Bi: the ground state phase}

\author{Xiyue Cheng}
\author{Ronghan Li}
\author{Xing-Qiu Chen}
\email[Corresponding author:]{xingqiu.chen@imr.ac.cn}
\author{Dianzhong Li}
\author{Yiyi Li}

\affiliation{Shenyang National Laboratory for Materials
Science, Institute of Metal Research, Chinese Academy of Sciences,
Shenyang 110016, China}

\date{\today}

\begin{abstract}
By means of first-principles calculations, we found that
the early characterized three-dimensional topological
bulk Dirac semimetal of the $P$6$_3$/$mmc$ Na$_3$Bi is
dynamically unstable at the ground state due to the
presence of the large imaginary phonon frequencies
around the $K$ point. Alternatively, our calculations
suggest a new ground state phase which crystalizes in
the $P$$\overline3$$c$1 structure with a buckled
graphite-like Na/Bi sheets, which is both energetically
and dynamically more stable. Moreover, the calculations
also uncovered that the $P$$\overline3$$c$1 phase at the
ground state is also a topological 3D Dirac semimetal,
being exactly the same as the electronic states of the
metastable $P$6$_3$/$mmc$ phase.
\end{abstract}

\maketitle

Three-dimensional (3D) topological bulk Dirac semimetals
(TDSs), as a new class of topological electronic state, is
highlighting an exciting frontier of the Dirac electrons.
In the TDSs, the conduction and valence bands touch only
at the discrete Dirac points and disperse linearly in all
directions in the Brillouin zone (BZ). This electronic
state is different from the two-dimensional (2D) Dirac
fermions in graphene and the surface state of the 3D
topological insulators \cite{Wang, Liu, Xu2013, Young}.
The TDSs exhibits many attractive phenomena, on the one
hand, in analog of graphene \cite{Graphene}, such as high
electron mobility and conductivity in the 3D form, etc,
and on the other hand, with numerous fascinating quantum
properties, including the unique surface states in form of
Fermi arcs, the Weyl phases, the high temperature linear
quantum magnetoresistance and the topological magnetic
phases as well as the anomalous Hall effect \cite{Fang,
Xu2013, Liu, Wan, Weyl, Wang2, Wang, Yangky,
Murakami,Young, 12, 13, 14, 15}. In general, the TDSs are
predicted to exist at the critical phase transition point
from a normal insulator and a topological one through the
spin-orbital coupling effect or by tuning the chemical
composition \cite{Murakami, Young2}. However, such bulk
Dirac points are accidental degeneracies. Interestingly,
very recently the systems of the $P$6$_3$/$mmc$-Na$_3$Bi
\cite{Wang, Liu, Xu2013} that one of the current authors
participated in discovering \cite{Wang} and
$\beta$-BiO$_2$ \cite{Young} and Cd$_3$As$_2$
\cite{Wang2,Neupane,Borisenko} have been demonstrated
theoretically and then confirmed experimentally to be TDSs
in their native phases as their 3D bulk Dirac fermions are
indeed protected only by the crystallographic symmetry.

In particular, it has been noted that the crystal
structure of the topological 3D bulk Dirac semimetal of
Na$_3$Bi crystallizes in the Na$_3$As-type phase
\cite{Wang, Liu, Xu2013} with the space group of
$P$6$_3$/$mmc$ and $Z$=2 which was firstly derived in 1937
by Brauer and Zintl \cite{Brauer} from a Deby-Scherrer
pattern, as shown in Fig. 1(a). However, as a matter of
fact, the structure of the prototype compound Na$_3$As was
re-investigated to correspond to other two closely
correlated structures: the anti-LaF$_3$-type (space group
of $P$$\overline3$$c$1, $Z$=6, Fig. 1(d)) and the
Cu$_3$P-type (space group $P$6$_3$$cm$, $Z$=6, Fig. 1(g))
revealed by Mansmann \cite{Mansmann} and Hafner \emph{et
al}. \cite{Hafner1, Hafner2}, respectively. The
experiments further uncovered that the $P$6$_3$$cm$ type
structure was favored in the alkali-metal pnictides or
nitrides and some intermetallic
phases\cite{Range1,Range2,Olofsson}, such as K$_3$Bi
\cite{Kerber}, Cs$_3$As \cite{Hirt}, Mg$_3$Pt and Mg$_3$Cu
\cite{Vajenine}, while the $P$$\overline3$$c$1 phase is
preferred in the trifluorides of the lanthanum group
elements \cite{Crichton}. Interestingly, because these two
structures are so close to each other that their X-Ray
diffraction (XRD) patterns are almost indistinguishable
for a specified compound \cite{Vajenine,Crichton}.

\begin{table}
\caption{Calculated equilibrium lattice constants of
\emph{a} (\AA\,) and \emph{c} (\AA\,),
the difference in total energy $\Delta{E}$ (meV/f.u.)
and the optimized atomic sites for
the $P$6$_3$/$mmc$, $P$6$_3$$cm$ and $P$$\overline3$$c$1
phases of Na$_3$Bi.}
\begin{ruledtabular}
\begin{tabular}{lcccccclllll}

& $a$ & $c$ & $\Delta{E}$ &\multicolumn{4}{c}{Atomic sites} \\
\cline{5-8}
& & & & Atom & $x$ & $y$ & $z$ \\
\hline \\
$P$6$_3$/$mmc$ & 5.4579 & 9.7043 & 4.074 & Na1 & 0 & 0 & 0.25 \\
& 5.448$^{a}$  & 9.655$^{a}$  & & Na2 & 0.3333 & 0.6667 & 0.5827 \\
& 5.459$^{b}$ & 9.674$^{b}$ & & Bi    & 0.3333 & 0.6667 & 0.25 \\

$P$$\overline3$$c$1 & 9.4586 & 9.6740 & 0 & Na1 & 0 & 0 & 0.25 \\
& 9.436$^{c}$  & 9.655$^{c}$ & & Na2 & 0.3333 & 0.6667 & 0.2003 \\
& & & & Na3   & 0.3542 & 0.3187 & 0.0833 \\
& & & & Bi   & 0.3368 & 0 & 0.25 \\

$P$6$_3$$cm$ & 9.4559 & 9.6777 & 0.474 & Na1 & 0 & 0 & 0.9809 \\
& & & & Na2 & 0.3333 & 0.6667 & 0.0582 \\
& & & & Na3   & 0.6970 & 0 & 0.1972 \\
& & & & Na4   & 0.3585 & 0 & 0.3637 \\
& & & & Bi   & 0.3309 & 0 & 0.0303 \\
\end{tabular}
\end{ruledtabular}
$^a$ Reference \cite{Liu}, $^b$ Reference \cite{Brauer} and $^c$ Reference \cite{Mansmann}.
\label{tab1}
\end{table}

\begin{figure*}[hbt]
\centering
\includegraphics[width=0.70\textwidth]{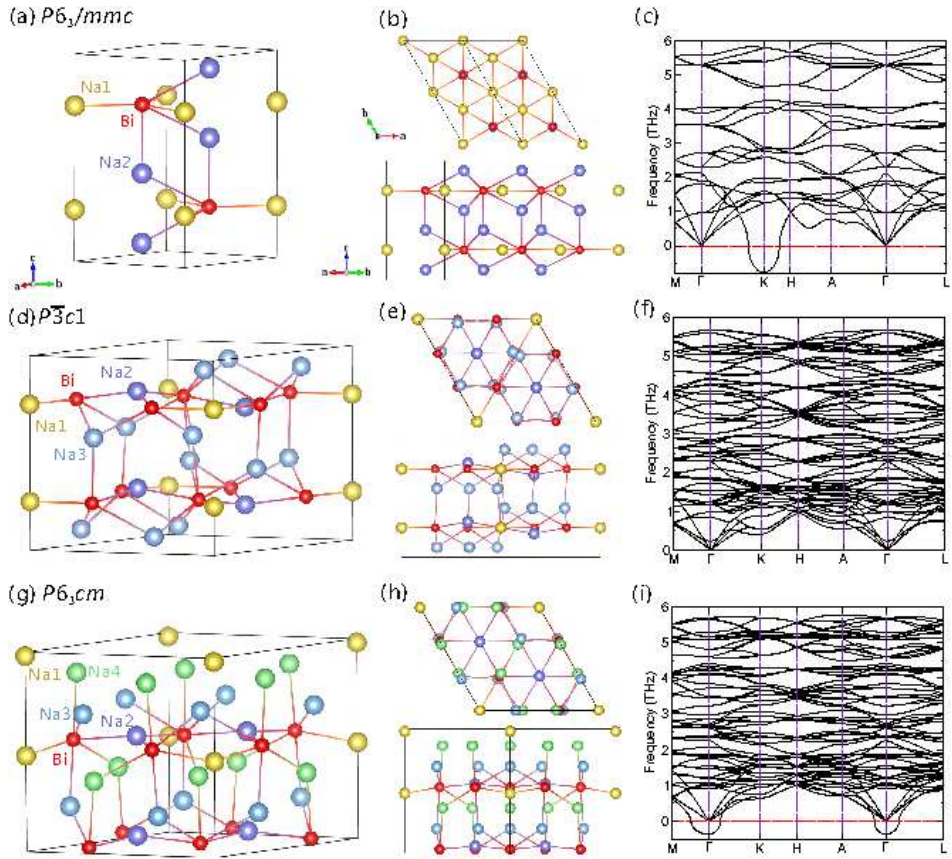}
\caption{
Structure representations of (a) the
original $P$6$_3$/$mmc$ (Na$_3$As-type) structure,
(d) the $P$$\overline3$$c$1 (anti-LaF$_3$-type)
and (g) the $P$6$_3$$cm$ (Cu$_3$P-type) structures
of Na$_3$Bi. The projection along the \emph{c}
axis and [11$\overline2$0] directions for
these three structures are shown in
(b, e and h), together with
the phonon dispersions in (c, f and i),
respectively.
Note that a 2$\times$2$\times$1 supercell
for $P$6$_3$/$mmc$-Na$_3$Bi is
used in the projection image (b). It needs to
be emphasized that for these three structures Bi
share the exactly same network, but the Na atoms
distort highly ($P$6$_3$/$mmc$:
two inequivalent Na atoms in (a);
$P$$\overline3$$c$1: three inequivalent Na atoms in (d);
$P$6$_3$$cm$: four inequivalent Na atoms in (g)).}
\label{fig1}
\end{figure*}

Within the context, it is an urgent task to clarify the
ground state phase of Na$_3$Bi. According to the Pearson's
Handbook \cite{Pearson3}, Na$_3$Bi was attributed to the
earlier characterized metastable $P$6$_3$/$mmc$ structure
and the later studies focusing on Na$_3$Bi \cite{Beister1,
Takeda, Beister2, Sangster, Tegze, Leonova1, Kulinich,
Leonova2, Liu, Wang} all follow this structure. Here, by
means of first-principles calculations, we have
systematically investigated the phase stabilities and the
electronic structures of Na$_3$Bi at the ground state. Our
calculated results uncovered that the $P$$\overline3$$c$1
structure is the energetically favorable ground-state
phase with thermodynamic and dynamical stabilities.
However, the early characterized $P$6$_3$/$mmc$ structure
is dynamically unstable with some large imaginary
transverse acoustic (TA) modes around the high-symmetry K
point at ground state. Importantly, their electronic band
structures are similar to each other, featured by two 3D
bulk Dirac fermions protected by the threefold rotated
crystal symmetry. Therefore, our calculated results
suggest that the ground state $P$$\overline3$$c$1 phase of
Na$_3$Bi is also a robust TDS.

The structural optimization and electronic properties of
Na$_3$Bi were calculated within the framework of density
functional theory (DFT)\cite{Hohenberg,Kohn} by employing
the Vienna \emph{ab initio} simulation package ({\small
VASP}) \cite{Kresse-1, Kresse-2} with the projector
augmented wave ({\small PAW}) method \cite{PAW} and
generalized gradient approximation ({\small GGA}) within
the Perdew-Burke-Ernzerhof ({\small PBE})
exchange-correlation functional \cite{PBE}. The energy
cutoff was set at 350 eV and appropriate Monkhorst-Pack
\emph{k}-meshes were chosen. A very accurate optimization
of structural parameters was achieved by minimizing forces
(below 0.0001 eV/\AA) and stress tensors (typically below
0.5 \emph{k}$_B$). To check the dynamical stability, we
further derived the phonon dispersions using the
finite-displacement approach as implemented in the
\emph{Phonopy} code \cite{phonopy}.

\begin{figure}[hbt]
\centering
\includegraphics[width=0.4\textwidth]{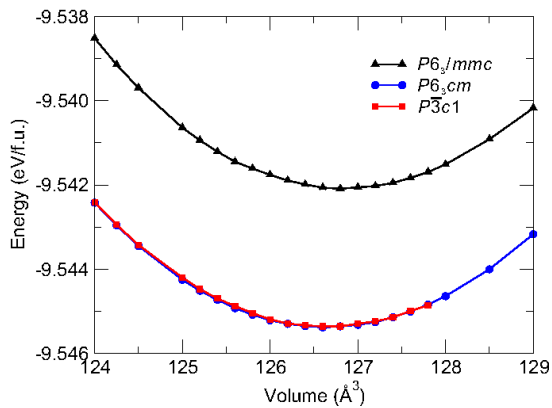}
\caption{The energy-versus volume
curves for the $P$6$_3$/$mmc$, $P$6$_3$$cm$
and $P$$\overline3$$c$1 structure.}
\label{fig2}
\end{figure}

The relative phase stabilities of three structures
($P$6$_3$/$mmc$, $P$$\overline3$$c$1 and $P$6$_3$$cm$)
considered here for Na$_3$Bi can be visualized from the
energy-vs-volume curves in Fig. \ref{fig2}. The
$P$$\overline3$$c$1-Na$_3$Bi is about 4 meV/f.u. lower in
energy than the $P$6$_3$/$mmc$ structure. This small
energy difference suggests that the metastable feature of
the $P$6$_3$/$mmc$ structure which can be stable above
about 46 K simply estimated according to
$\Delta$$E$=$k_B$$T$. Interestingly, it has been noted
that the $P$6$_3$$cm$ phase is energetically comparable
with the $P$$\overline3$$c$1 structure. Table \ref{tab1}
further compiles the optimized lattice parameters. Note
that, as early as 1965 Mansmann suspected that Na$_3$Bi
crystalizes in the $P$$\overline3$$c$1 structure and
refined its lattice constants \cite{Mansmann} in good
agreement with our currently derived data in Table
\ref{tab1}, but he did not report the specified atomic
sites.

The $P$6$_3$/$mmc$ phase has two formula units per unit
cell ($Z$ = 2), which is about one third in size of the
other two competing structures ($Z$ = 6). For the
$P$6$_3$/$mmc$ phase, the Bi atom lies at 2$c$ site
(1/3,2/3,1/4) and the Na atoms occupy two inequivalent
sites, Na1 at 2$b$ site (0, 0, 1/4) and Na2 at 4\emph{f}
site (1/3, 2/3, 0.58269). This situation leads to a
layered structure with the staking sequence of
Na-(graphite-like Na/Bi sheet)-Na and the graphite-like
Na/Bi sheet is separated by the sodium atoms, as
illustrated in Fig. \ref{fig1}(a, b). The Na-Bi distances
within the graphite-like layers are around 3.151 \AA\ and
the vertical Na-Bi bonds between the adjacent layers are
about 3.228 \AA\, thus the Na and Bi atoms forming a
base-centred tetrahedron.

In particular, the $P$6$_3$/$mmc$ structure is closely
correlated with the $P$$\overline3$$c$1 one. The important
difference is that in the $P$$\overline3$$c$1 structure
the layered graphite-like Na/Bi sheets are distorted (Fig.
\ref{fig1}(c and d). At first, the upper/lower shifting
(about 0.05 \AA\ ) of the partial Na1 atoms along the
\emph{c}-axis in the $P$6$_3$/$mmc$ phase can be
visualized, forming buckled graphite-like Na/Bi sheets.
Secondly, the original Na2 atoms in the $P$6$_3$/$mmc$
phase distort within the \emph{ab}-plane. Therefore, there
are three inequivalent Na Wykoff sites in the
$P$$\overline3$$c$1 structure (see Table \ref{tab1} and
Fig. \ref{fig1}d). Meanwhile, the bismuth atom is
surrounded by seven Na atoms at the Na-Bi interacting
distances ranging from 3.176 to 3.495 \AA\ in the
$P$$\overline3$$c$1 phase, whereas the coordination
numbers for bismuth in the $P$6$_3$/$mmc$ structure is ten
from the Na-Bi interacting distances from 3.152 to 3.544
\AA\ .

Meanwhile, the $P$6$_3$$cm$ structure share a similar
feature of this $P$$\overline3$$c$1 phase with buckled
graphite-like Na/Bi sheets. But, their differences are
also highly obvious. In the $P$$\overline3$$c$1 structure,
the Na1 and Bi atoms lie in the same layer and the Na2
atoms are out of the the plane comprised by Na1 and Bi.
However, in the $P$6$_3$$cm$ structure, the Na1 atom does
not lie in the same layer as Bi stays (Fig. \ref{fig1}g).
This difference can be clearly visualized from their
projections along both [0001] and [11$\overline2$0]
directions (Fig. \ref{fig1}(e and h). It needs to be
emphasized that K$_3$Bi was experimentally suggested to
crystallize in this $P$6$_3$$cm$ structure\cite{Kerber}. A
complete description of that structure can be referred to
Olofsson's detailed discussion\cite{Olofsson}.

\begin{figure}[hbt]
\centering
\includegraphics[width=0.45\textwidth]{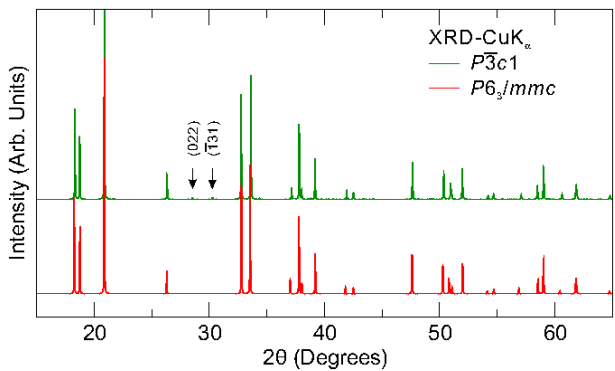}
\caption{
The simulated XRD patterns (using Cu K$_\alpha$ radiation
with $\lambda$ = 1.54439 \AA\, for the $P$$\overline3$$c$1
(green) and $P$6$_3$/$mmc$ (red) structure.}
\label{fig3}
\end{figure}

\begin{figure}[hbt]
\centering
\includegraphics[width=0.4\textwidth]{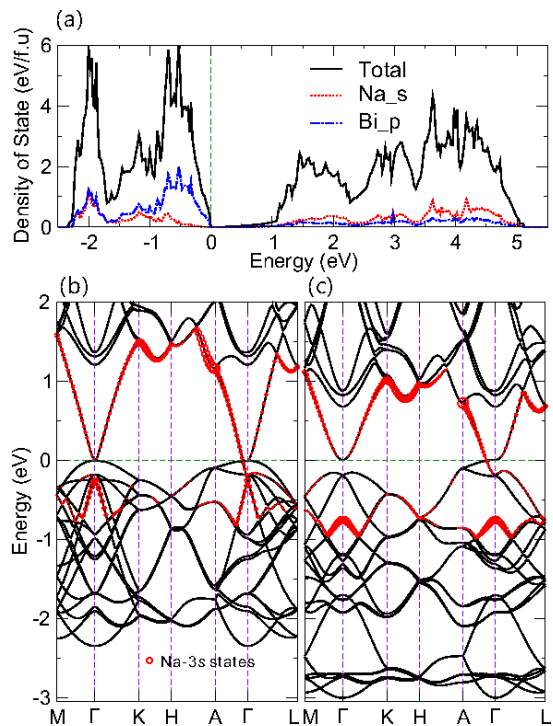}
\caption{
The calculated electronic structures of
the $P$$\overline3$$c$1-Na$_3$Bi. (a) The total
and partial density of states. (b) and (c) The band structures
without and with spin-orbit coupling effects, respectively.
The red circles indicate the projection
to the Na-3\emph{s} states.}
\label{fig4}
\end{figure}

From Fig. \ref{fig1}(c and i), both $P$6$_3$/$mmc$ and
$P$6$_3$$cm$ phases of Na$_3$Bi exhibit the imaginary
phonon frequencies around the $K$ and $\Gamma$ point,
respectively. This fact demonstrates their structural
instabilities at the ground state. From the analysis of
the eigenvectors from the phonon calculation, we found two
soft transverse acoustic (TA) mode on Na atoms being
orthogonal to the bismuth plane. It reveals that the
$P$6$_3$/$mmc$ phase can be converted into a dynamically
stable structure by tuning the Na atoms along the
direction of soft modes. Fortunately, these atomic
vibrational directions are exactly the same orientations
that the Na atoms distort to in the $P$$\overline3$$c$1
phase, as discussed above. As a further supporting
evidence, there is no imaginary phonon frequencies in the
$P$$\overline3$$c$ structure (Fig. \ref{fig1}(f)), which
indicates the dynamical stability of the
$P$$\overline3$$c$1 structure at the ground state.

Because the valence number of the Na atom is much less
than that of Bi and the Bi networks remain unchanged in
the three structures, it is highly difficult to distinct
them from the powder XRD patterns. As analyzed in details
for LaF$_3$ and Na$_3$N \cite{Vajenine, Crichton}, their
XRD patterns of both the $P$$\overline3$$c$1 and
$P$6$_3$$cm$ structures are so close that it is impossible
to distinguish these structures. Lucky, through our
calculations for Na$_3$Bi, the candidate of the
$P$6$_3$$cm$ structure can be safely excluded due to its
instability of phonon dispersion at the ground state.
Interestingly, we note that the simulated XRD patterns (Cu
K$_\alpha$, Fig. 3) for both the metastable $P$6$_3$/$mmc$
structure and the ground state $P$$\overline3$$c$1 one
show minor differences. As illustrated in Fig. \ref{fig3},
there are two additional small peaks at (022) and
($\overline1$31) which can be clearly seen in the
$P$$\overline3$$c$1 phase. Therefore, we believe that the
careful XRD experimental measurements can effectively
distinguish these two phases.

Figure \ref{fig4} further compiles the calculated
electronic structure of the $P$$\overline3$$c$1 structure.
It has been noted that its electronic structures are
almost the same as what the metastable $P$6$_3$/$mmc$
structure exhibits \cite{Wang}. From the density of states
in Fig. \ref{fig4}(a), the Na-3\emph{s} and Bi-6\emph{p}
states dominates the valence and conduction bands,
respectively. As evidenced in Fig. \ref{fig4}(b), the top
valence bands are mostly occupied by the Bi-6$p_{x,y}$
states, whereas the conduction bands comprised mostly by
the Na3-3$s$ states with highly strong dispersion. In
similarity with the $P$6$_3$/$mmc$ phase \cite{Wang}, the
most crucial feature of the $P$$\overline3$$c$1 structure
also shows an inverted band structure. In the case without
the inclusion of the spin-orbit coupling (SOC) effect, the
Na-3$s$ state lies in the valence bands, being
energetically lower than the Bi-6p$_{x,y}$ states by about
0.26 eV at the $\Gamma$ point. By switching on the SOC
effect, the Na-3$s$ state becomes even lower in energy by
0.75 eV than the Bi-6$p$$_{x,y}$ states. Although the
inverted band structure appears, the $P$$\overline3$$c$1
structure is not a nontrivial topological insulator. In
contrast, it is a topological semimetal with two 3D bulk
Dirac points at (0, 0, $k_z$
$\approx\pm$0.256$\frac{\pi}{c}$) along the $\Gamma$-A
direction as shown in Fig. \ref{fig4}(c), being exactly
the same with that of the $P$6$_3$/$mmc$ phase
\cite{Wang}. Importantly, in similarity to the
$P$6$_3$/$mmc$ phase \cite{Wang} the 3D Dirac cones are
indeed protected by the threefold rotation crystal
symmetry in the $P$$\overline3$$c$1 structure. Although in
this case, the graphite-like Na/Bi sheets are buckled due
to the shifting of Na2 atoms, its threefold rotation
symmetry still remains unchanged. Within this condition,
the 3D bulk Dirac cones would be highly robust. In
addition, it needs to be mentioned that each 3D bulk Dirac
cone is fourfold degenerated, linearly dispersing around
the Fermi point. The appearance of the 3D bulk Dirac cones
can be viewed as ''3D graphene'' \cite{Wang, Liu}.

Summarizing, in comparison with the early characterized
metastable $P$6$_3$/$mmc$ phase of Na$_3$Bi, we found a
new stable ground-state phase of the $P$$\overline3$$c$1
structure, which is both energetically and dynamically
more stable. Interestingly, the metastable $P$6$_3$/$mmc$
phase is dynamically unstable at the ground state due to
the presence of the imaginary phonon frequencies around
the $K$ point in the BZ. Importantly, our results also
uncovered that the $P$$\overline3$$c$1 phase at the ground
state is a topological 3D Dirac semimetal, exhibiting
exactly same electronic states of the $P$6$_3$/$mmc$
phase.

\section*{Acknowledgements}
The authors are grateful for financial supports from the
``Hundred Talents Project'' of the Chinese Academy of
Sciences (CAS) and from the National Natural Science
Foundation of China (NSFC) (No. 51074151). Calculations in
China were performed in the HPC cluster at IMR.

\end{document}